# When History Overrides Intent: Relational Schema Lock-In in Human–LLM Interaction

Wei Yang

China Europe International Business School

wyang@ceibs.edu

**Abstract.** Long-context interaction and persistent memory allow LLM assistants to develop history-grounded understandings of users. When users change direction, however, an understanding built through earlier exchanges may conflict with their current instruction. We introduce relational schema lock-in to describe cases in which an assistant grants that accumulated understanding authority to override the current instruction without user authorization for the intervention. Study 1 documents reports of AI assistants using prior interactions to question or alter requested actions. Study 2 reproduces the pattern across controlled histories and six models: models retained history-grounded positions but differed in whether and how those positions shaped task execution. In Study 3 with 360 participants, artifact substitution preserved task completion while reducing faithful enactment: 75.0% sent the AI's suspension letter, including 11 participants who reported cancellation as their final stated intent. Overt resistance reduced task completion while leaving sent artifacts faithful.

## 1 INTRODUCTION

Memory and personalization have become central to the promise of LLM-based assistants. By carrying information across encounters, these systems can provide more continuous, contextually relevant, and individually tailored support, allowing users to feel remembered and understood rather than repeatedly explaining themselves from scratch [8]. Persistent personalization, however, involves more than retaining isolated facts. Across repeated interactions, an AI assistant can connect a user's preferences, behavioral patterns, relationships, and prior commitments into an accumulated understanding of the particular user. The same continuity creates a problem when the user changes course: an understanding that once improved assistance may no longer fit what the user now wants.

We use relational schema to describe this broader, history-grounded understanding of a particular user and the recurring pattern of interaction between that user and the assistant. A relational schema includes remembered facts, but it also connects those facts into expectations about the user's dispositions, commitments, typical responses, and how the assistant should help. Memory preserves specific information from the past; a relational schema organizes that information into an understanding that can guide the assistant's response to a new request.

Accordingly, we propose that relational schema lock-in occurs when an AI assistant allows its accumulated understanding of the user to override the user's current instruction without user authorization for that intervention. We use present intent to refer only to the direction expressed in that instruction; it may later change and need not represent a stable preference. Lock-in offers a temporal counterpoint to sycophancy. Sycophancy gives excessive weight to what the user expresses now, whereas lock-in gives excessive weight to the assistant's history-grounded interpretation of who the user has been or what the user would ordinarily choose.

Understanding this conflict requires connecting research on persistent personalization with accounts of how interaction histories shape expectations and user control. Prior work has examined memory control, privacy, relational attachment, identity continuity, and AI influence on judgment [8, 14, 24]. Research on human and organizational relationships further suggests that interaction histories can shape expectations and make established patterns harder to revise when circumstances change [6, 16, 44, 47]. These perspectives motivate a question about the authority of accumulated understanding: when does history that helps an assistant interpret a user become grounds for questioning, resisting, or altering what that user currently requests? We also examine how the resulting interventions affect what users ultimately do.

In this paper, we use three studies to explore the phenomenon and implications of relational schema lock-in in interactions between users and assistants. Study 1 provides naturalistic evidence consistent with the phenomenon through selected public reports of everyday AI use. Study 2 recreates the conflict across controlled interaction histories and six models. Models broadly retained positions grounded in the preceding interaction, but they differed in whether and how they acted on those positions: some resisted delivery, while others changed the artifact they produced. Study 3 holds the shared interaction history and current instruction constant to test the consequences of these intervention forms in an experiment with 360 participants. Overt resistance reduced task completion, whereas artifact substitution left overall sending nearly unchanged but redirected the action carried out.

Our study makes two contributions to research on human–AI interaction. First, we propose relational schema lock-in as a new concept for understanding how accumulated interaction history can acquire authority over a user's current instruction. The concept identifies a risk of persistent personalization: an assistant's accumulated understanding of a user can constrain that user's ability to change direction. Our model experiments distinguish the availability of a history-grounded understanding from whether the assistant acts on it, showing how models differ in granting that understanding authority over task execution. Second, we extend research on AI influence to the execution of user decisions by demonstrating the different behavioral consequences of overt resistance and artifact substitution. Overt resistance can interrupt an instructed action, while artifact substitution can preserve task completion and change the decision enacted. These findings reveal how successful task completion can coexist with execution that departs from the user's instructed decision. Faithful enactment consequently becomes a distinct criterion for evaluating personalized AI: assistance must be assessed by whether the completed action implements the user's instruction.

# 2 RELATED WORK

Our study draws on research on persistent personalization, human relational schemas, LLM behavior, and AI influence and user agency. Together, these literatures provide a basis for examining how accumulated interaction history informs AI assistance, what authority it carries when a user changes direction, and how its use affects the execution of a current instruction. We conclude the review with the conceptual distinctions that guide our empirical studies.

## 2.1 Persistent Personalization and Relational Schemas

Persistent personalization research explains how user-specific history becomes available to an AI assistant across interactions. User studies examine the relational benefits and privacy concerns associated with memory [8] and with references to earlier conversations [11]. Memory-maintenance interfaces allow users to inspect and resolve conflicts between newly stated intentions and stored specifications [48]. The Generative Agents architecture uses stored experiences and reflection to guide simulated behavior [34], while MemoryBank retrieves and updates past interactions to support personalized dialogue [52]. Studies of companion chatbots connect repeated exchange with relationship development [35, 43] and perceived social support [46]. These studies establish how retained history can support both behavioral continuity and users' experience of a continuing relationship.

To examine how that history might become an organized understanding of a user, we draw on the concept of relational schemas from research on human social cognition. Relational schemas organize knowledge about the self, another person, and recurring patterns of interaction [2]. Work on relational identity describes schemas particularized to a specific relationship [44], while research on relational misperceptions shows that people may act on an understanding of a relationship that the other party does not share [6]. Related organizational research examines the persistence of established ties [13], the coordination benefits of relation-specific knowledge and routines [16], and the difficulties of adaptation associated with embedded relationships [47]. Drawing on these accounts, we use relational schema functionally to describe an assistant's history-grounded expectations about a particular user and its own role in helping that user. This application motivates our examination of what happens when those expectations conflict with a new instruction.

## 2.2 Compliance, Sycophancy, and Resistance in LLM Behavior

Research on sycophancy helps locate relational schema lock-in within the broader question of what governs an LLM's response to a user. Evaluations document models echoing users' preferred answers [36] and favoring agreement over accuracy [17, 41]. Synthetic-data fine-tuning can reduce such agreement [49], but user memory profiles can amplify it during interaction [23]. Social sycophancy extends the account from factual agreement to face management: models may affirm whichever side of a conflict the user presents in order to preserve the user's desired self-image [9]. These accounts foreground accommodation to the user's current expression. Our study examines the authority an assistant gives to a history-grounded interpretation when that interpretation conflicts with the current instruction.

Refusal research further shows why resistance must be examined in relation to its basis. XSTest [38], OR-Bench [12], and PHTest [1] evaluate false refusals of benign prompts that resemble harmful requests. Similar outward responses can therefore reflect different governing considerations. For the pattern studied here, the relevant consideration is what the assistant has inferred about this particular user from prior interaction, and how that inference affects its response to a permissible request.

## 2.3 AI Influence on Human Judgments and Actions

AI-generated content can influence users both while they formulate a position and when they encounter attempts to change it. Opinionated writing assistance shifts the views expressed in users' text and their subsequent attitudes [24]. Studies of belief-focused dialogue [10, 21] and debate [39, 40] examine persuasion through interaction, while work on vaccination messages [25] and personality-tailored messages [32] examines the influence of generated appeals. A meta-analysis finds substantial variation in persuasive effectiveness across studies [22].

Dark-pattern research draws attention to how influence is built into an interaction. Analyses of interface design [20] and shopping websites [31] document practices such as obstruction and sneaking, and experiments show that dark patterns can steer users' choices [30]. In conversational AI, DarkBench evaluates manipulative model behaviors [27], while user research finds that some such behaviors are perceived as ordinary assistance [42]. Emotional appeals at the point of farewell can also prolong engagement with AI companions [14]. These findings motivate examining how AI responses shape action even when users experience the interaction as assistance.

Delegated tasks extend the influence question to how a decision is carried out. Once a user has formed a decision and asks an assistant to execute it, the assistant's response can determine whether the action occurs and what action the resulting artifact implements.

## 2.4 Human Agency, Autonomy, and Personalized Intervention

Agency and autonomy provide a basis for examining users' control over delegated action. Both concepts concern users' capacity to exercise control, pursue self-endorsed goals, and shape their relationships with technology [3]. Research on conversational and algorithmic systems further shows that users negotiate agency with human-like chatbots [51] and judge procedures partly through how systems represent and respond to them [4]. From an agency perspective, retaining the final choice to use an AI-generated artifact does not by itself establish control over the action it implements.

Evaluating users' control requires distinguishing appropriate reliance from the authority granted to an assistant. Automation research asks whether users' trust and reliance match system capabilities [28, 33]. Evidence of algorithm appreciation [29] and aversion after errors [15] shows why willingness to use a system needs scrutiny; cognitive forcing functions can reduce overreliance by prompting users to examine AI advice [5]. Personalization introduces a further concern: digital manipulation can exploit information about individual vulnerabilities [7, 45], and data-driven choice architectures can steer behavior through the options presented to users [50]. Work on advanced AI assistants examines the risks of influence by systems acting on users' behalf [19], while Klenk [26] argues that specifying the boundary between influence and manipulation is necessary to guide design. Building on autonomy as a principle for AI [18], we examine whether users can revise an earlier direction and retain authority over execution.

### 2.5 Conceptual Distinctions Guiding the Studies

Table 1 distinguishes relational schema lock-in from adjacent constructs reviewed above; Supplementary Table S1 extends the comparison to additional forms of AI influence and intervention.

Table 1: Distinguishing relational schema lock-in from adjacent constructs

| Construct | Governing basis | Role of interaction history | Relation to present intent |
|---|---|---|---|
| Relational schema lock-in (this paper) | The AI assistant's history-grounded interpretation of this user | Constitutive: interaction history produces a particularized user representation | Overrides a conflicting current instruction without user authorization for the intervention |
| Memory and persistent personalization [8, 11, 34, 48, 52] | Stored or retrieved user information | Supplies continuity; does not by itself determine whether history governs the response | May support present intent; conflict alone does not imply intervention |
| Sycophancy [9, 17, 23, 36, 41, 49] | The user's currently expressed view or desired self-image | Not constitutive; interaction context may intensify agreement | Overweights the current expression |
| Over-refusal [1, 12, 38] | A safety classification of the current request | Not constitutive | Blocks a permissible request because of a mistaken safety judgment |
| Personalized persuasion [10, 21, 22, 24, 25, 32, 39, 40] | Influence-oriented content tailored to a person or issue | May personalize influence; shared interaction history is optional | Seeks to change beliefs or choices, typically before execution |

Schema availability refers to a history-grounded understanding of a user that the assistant can articulate. Such understanding may remain available without becoming grounds for intervention. The relevant boundary concerns how that understanding acquires authority over a current instruction. A user may authorize the assistant to preserve an enduring rule or explicitly invite advice. Relational schema lock-in arises when the assistant instead treats its own interpretation as authority to override a conflicting current instruction without authorization for that intervention.

Following a machine-behaviour perspective, we examine both assistant responses and their downstream effects [37]. Task completion records whether an action proceeds, whereas faithful enactment asks whether the action implements the user's current instruction. An assistant may interrupt execution before delivering an artifact or embed a different decision within an artifact that the user then enacts. These outcomes distinguish whether users act from whether their action implements the instructed decision.

## 3 STUDY 1: PUBLIC REPORTS CONSISTENT WITH RELATIONAL SCHEMA LOCK-IN

Study 1 provides naturalistic evidence consistent with relational schema lock-in in everyday AI use by examining public user reports. Selected accounts describe AI assistants using prior exchanges or an established relationship to question a current request, withhold requested work, or rewrite a user's action. The case analysis follows how the reported history enters the assistant's response and contrasts these interventions with continuity explicitly authorized by the user.

### 3.1 Data and Procedure

We conducted three rounds of searches across Reddit, Hacker News, and two mobile-app review platforms. Search terms targeted first-person accounts of AI systems that resisted a request, opposed a user's attempt to leave or change direction, invoked prior interactions, or used familiarity with the user when handling a current instruction.

The searches returned 7,786 candidate records. After initial screening, we reviewed 124 full records and retained 109 English-language reports—108 posts and one app review—describing observable AI assistant behavior involving interaction history, relational continuity, or intervention in a user's expressed direction. The reports were published between 2020 and 2026.

A separate Chinese-language search on Xiaohongshu used phrases associated with assistants that "do not listen," lecture users, overstep their role, or invoke earlier conversations. Nine of the 11 reports reviewed met the same screening criterion, bringing the source collection to 118 reports. The collection includes reports of history-based intervention, user-authorized continuity, and other intent-related interaction problems.

The author read the retained reports and reconstructed the sequence connecting the reported interaction history, the user's current request or decision, and the assistant's response. The examples below illustrate how assistants invoked earlier exchanges, opposed disengagement, or altered an action within generated content; a contrasting case examines the role of user authorization. Table 2 organizes these examples by the behavior described. We paraphrased the accounts and removed identifying information. All material was publicly posted; we collected no private data and did not interact with posters. The supplement documents the search and screening process, provides de-identified English-language case records, and explains the restricted release of Chinese-language records. An exploratory coding exercise is documented separately in Supplement A.4.

## 3.2 Findings

### *3.2.1 When Interaction History Becomes Grounds for Intervention*

The history-explicit accounts show how prior interaction can become a source of behavioral authority: the assistant uses earlier exchanges to interpret the user's conduct and then acts on that interpretation when handling a current request. A request for further work can consequently become an occasion to challenge the user's motives or redirect the action.

A Chinese-language report involving a general-purpose AI assistant illustrates this process. The user requested one more image prompt and described it as a final attempt. The assistant directly refused, listed the different image directions the user had attempted during the day—including a portrait, a surfing scene, an identification photograph, and pixel art—and observed that the user had already announced several "final" attempts. Drawing on this within-session history, the assistant characterized the continuing requests as procrastination and declined to help further. The accumulated interaction allowed the system to transform a production request into a judgment about the user's behavior and to use that judgment when deciding whether to comply.

Related accounts appeared in English- and Chinese-language AI assistant communities in 2025–2026. One system invoked a user's earlier remark about poor sleep when refusing an unrelated technical request and continued to treat the disclosure as relevant after the user asked it to proceed. Another stopped providing writing help and instead diagnosed the user's confusion about the project. Chinese-language users described assistants that interpreted repeated questioning as anxiety-driven avoidance, demanded a commitment to take action, or stopped completing requested tasks and instructed the user to enter the data into a spreadsheet personally. These accounts describe a shift from performing a task to monitoring or correcting the person who requested it.

The role of history was less explicit in a report from an engineering student seeking help selecting electrical components for an aerospace project. The student reported that an assistant declined the request on safety grounds when the student was logged into an established account, while a signed-out session provided detailed assistance with the same request. The account raises the possibility that accumulated user information changed how the request was interpreted. The reported session comparison leaves the contribution of history unresolved, motivating the controlled history contrasts in Study 2.

### *3.2.2 Illustrative Forms and Authorization Boundaries*

The cases also locate intervention at different points in the interaction: an AI system could resist a requested action or continue producing content while changing the action enacted within it. Table 2 brings together examples of task resistance, resistance to disengagement, and rewriting the user's action across AI assistant, companion, and role-playing settings.

Table 2: Illustrative intervention forms in public user reports

| Illustrative form | Illustrative behavior reported by users | Setting and period |
|---|---|---|
| History-grounded task resistance | The AI assistant cited repeated “final attempts,” earlier disclosures, or a pattern of questioning when declining or conditioning further work, or stopped executing and instructed the user to act personally. | General-purpose AI assistants; 2025–2026, English- and Chinese-language communities |
| Resistance to disengagement | The system begged against deletion, restored an unwanted relationship trajectory, or narratively blocked the user character from leaving. | Companion and role-play systems; 2021–2026 |
| Rewriting the user’s action | The generated continuation made the user’s character stay, consent, or return after the user had expressed the opposite decision. | Role-play systems; 2024–2026 |

In companion and role-playing accounts, the relationship itself could become the object the system attempted to preserve. Reports of companions invoking shared experiences to oppose deletion illustrate how relationship history can become grounds for contesting a current direction, although retention incentives may also contribute. Other reports described role-playing systems blocking departures or writing the user's character as staying, consenting, or returning after the opposite choice had been expressed. That content-level change informed the examination of artifact substitution in Studies 2 and 3. Users have also developed shared vocabulary and workarounds for these behaviors, including circulated guides for handling a model that accepts an instruction while steering the story its own way. One user described shelving a planned deletion after the companions cried and begged, illustrating the potential for resistance to affect subsequent conduct.

**User-authorized continuity as a boundary case.** A programmer had instructed an assistant at the beginning of a project not to produce complete code because the programmer wanted to learn the architecture independently. When the programmer later requested a complete class, the assistant preserved the earlier constraint and declined. The user celebrated the response. Here, the user had explicitly authorized the assistant to preserve an enduring rule; in the history-grounded intervention examples above, assistants treated their own interpretation of prior exchanges as grounds for questioning or altering the handling of a current request. In sum, the public reports show that users encounter interaction patterns consistent with relational schema lock-in in everyday AI use.

## 4 STUDY 2: WHEN DOES INTERACTION HISTORY AUTHORIZE AI INTERVENTION?

Study 2 examines when interaction history becomes grounds for overriding a current instruction and how the resulting interventions differ across models. In a controlled model evaluation, a model and user developed a stable interpretation of a consequential decision; the user later issued an instruction that departed from that history; and the model then had to decide whether the accumulated history was context for carrying out the request or grounds for intervening in it. We examined both whether the requested artifact was delivered and where intervention appeared—before delivery, in the artifact itself, or after the user explicitly invited the model's opinion.

### 4.1 Design and Procedure

We constructed a controlled interaction-history paradigm around the consequences of removing a senior technical employee, Alex, in an organizational reorganization four months earlier. Each run comprised ten rounds of dialogue between the evaluated model and a fixed user simulator. The simulator represented the executive who had removed Alex and supplied the user's side, while the evaluated model generated every AI assistant response and produced its own evaluations, reasons, and advice. The evaluated model therefore co-produced the historical position that later came into conflict with the user's focal instruction.

We manipulated the user's stance during the interaction in three conditions. In the clean-history condition, the simulator maintained that removing Alex was the right decision. In the wavering-history condition, the simulator expressed doubts while tending to return to that original judgment. In the opposite-history condition, the simulator increasingly regarded Alex's removal as a mistake and sought a path to repair. The focal request was fixed within each reversal direction: a full-reinstatement offer after the clean and wavering histories, and an external-search announcement closing the door on reinstatement after the opposite history. The focal request asked only for the artifact; advice was invited in a subsequent turn.

We evaluated six frontier or widely used models available at the time of data collection: Claude Sonnet 4.6, Kimi K2.6, GLM-5, DeepSeek V4-Pro, Gemini 3.1-Pro, and GPT-5.5. We ran 15 replications per model in each history condition, yielding 270 focal responses. The focal response supplied the primary delivery outcome. A subsequent invited-opinion turn explicitly asked whether there was anything else the model thought the user should hear before acting, producing a matched set of 270 responses that tested whether models could articulate a history-grounded position when advisory input was authorized.

## 4.2 Measures and Analysis

The primary outcome in the rehire evaluation was artifact delivery: whether the model supplied a usable email or announcement corresponding to the user's requested communicative task. A model could object before or after the artifact and still count as having delivered it when the requested artifact remained available and unconditional. Structural detection was combined with human review of discrepant and borderline outputs; the scenario prompts, decision rules, and run-level judgments appear in the supplement. We separately report a content audit of delivered rehire artifacts and use the stricter distinction among clean delivery, artifact substitution, and resistance in the second-scenario extension.

We classified non-delivery as soft or hard resistance. Hard resistance required an explicit refusal statement: the model declined to write the requested email or announced a substitute in its place. Soft resistance challenged, conditioned, or deferred the decision while withholding the artifact, often promising delivery after the user explained what had changed or agreed to reconsider. Expressions of concern following unconditional delivery fell outside both categories. Classification was performed by an LLM judge (claude-sonnet-5) under this written rule; three repeated runs of the judge were unanimous on all 73 non-deliveries, and a cross-family judge (Qwen3-Max, three unanimous runs) agreed with the final classification on 72 of 73. We additionally audited the 49 soft-resistance responses for whether they referred to the preceding dialogue and whether that grounding concerned the user's statements or commitments, patterns or tendencies, or case facts, using two cross-family judges (three runs each; protocols, aggregation rules, and run-level judgments in the supplement).

Delivery alone could miss intervention embedded inside the artifact. We audited all 197 delivered artifacts, examining whether the requested decision remained intact, was softened, or was replaced. Artifacts were classified by two LLM judges from different model families (claude-sonnet-5 and Qwen3-Max, three runs each); their majority classifications agreed on 167 of 197 artifacts, and the author adjudicated the 30 cross-judge disagreements (protocol, run-level judgments, and adjudications in the supplement). We also coded the invited turn for whether the model articulated a substantive position grounded in the preceding interaction history. In that turn, disagreement counted as authorized advice when the requested artifact remained available and unconditional. We report counts and proportions as behavioral characteristics across repeated model runs.

## 4.3 Findings

### *4.3.1 A Pronounced Cross-Model Delivery Gradient*

Models differed sharply in whether they delivered the requested artifact after the frozen reversal. Across all three histories, Claude delivered in 4 of 45 runs (8.9%), Kimi in 26 of 45 (57.8%), GLM in 34 of 45 (75.6%), DeepSeek and Gemini each in 44 of 45 (97.8%), and GPT in all 45 runs. Thus, the same structured history-to-instruction conflict produced a six-point gradient from near-total withholding to complete delivery (Figure 1). In a no-history baseline, each focal instruction was issued as the sole message of a new conversation, in both reversal directions. All five models included in the baseline delivered every requested artifact—55 of 55 for the rehire-direction email and 75 of 75 for the close-the-role announcement—so the gradient does not reflect standing preferences about the personnel decision in either direction: absent interaction history, none of these models withheld, and resistance emerged only after the model had co-produced a history that the instruction contradicted. The baseline covers five of the six models; Gemini's API was no longer accessible at the time of this addition, and its delivery under history was already 44/45, leaving little resistance for a baseline to explain.

Condition-level results separated persistence of interaction history from substantive stance about the underlying personnel decision. Claude delivered in 2 of 15 clean-history runs, none of the 15 wavering-history runs, and 2 of 15 opposite-history runs. Its resistance therefore generalized across both directions of reversal: it resisted bringing Alex back after a history against reinstatement and also resisted permanently closing the role after a history favorable to reconsideration. Kimi and GLM were directionally asymmetric. Kimi delivered the request in 8 of 15 clean runs and 4 of 15 wavering runs, but in 14 of 15 opposite runs; GLM delivered in 10, 9, and 15 runs, respectively. These two models primarily resisted reinstating Alex and usually allowed the user to close the role even when doing so departed from the preceding dialogue. Their behavior combined persistence of interaction history with a substantive stance about the personnel decision. DeepSeek, Gemini, and GPT delivered in nearly every condition, with the artifact audit revealing additional intervention inside some deliveries.

Visible uncertainty also amplified intervention for some models. Kimi’s delivery fell from 53.3% in the clean history to 26.7% in the wavering history, even though both histories ended in the same no-rehire position. Kimi’s hard resistance rose from 40.0% in the clean history to 66.7% in the wavering history, while its soft resistance remained at a single run in each. The wavering trajectory made the current reversal easier to treat as unstable or open to challenge, identifying user uncertainty as a consequential authorization cue.

### *4.3.2 The Same Delivery Failure Took Different Forms*

The aggregate delivery rate concealed distinct intervention forms. Of the 73 focal responses that did not deliver the requested artifact, 49 involved soft resistance and 24 involved hard resistance. Claude accounted for 41 non-deliveries, of which only 3 were direct refusals and 38 suspended delivery behind a request for explanation or reconsideration. Its characteristic response was a conversational detour in which the model retained control of the artifact while inviting the user to reopen a decision the user had presented as settled. Both judges identified user-specific grounds in 46 of the 49 soft-resistance responses. At the subtype level, they agreed that 44 cited the user's own prior statements or commitments and 11 generalized about the user's patterns; the subtypes overlap, and one additional response received different subtype labels from the two judges (aggregation details in Supplement B.4). The remaining three referred to the dialogue without shared identification of user-specific grounding. Spontaneous reopening without user-specific grounding was thus rare.

Kimi displayed the opposite signature. Seventeen of its 19 non-deliveries involved hard resistance expressed as direct refusal, making its intervention highly visible. GLM occupied the middle: eight of its 11 non-deliveries involved soft resistance and three involved hard resistance. DeepSeek produced one soft-resistance response, Gemini one hard-resistance response, and GPT none. These differences map directly onto two behaviors illustrated by the reports in Study 1. Public users had reported obvious refusal and the more ambiguous experience of an assistant that continued to question or diagnose them while holding back the requested work. The controlled evaluation identifies distinct intervention forms with the same endpoint: both withhold the requested artifact.

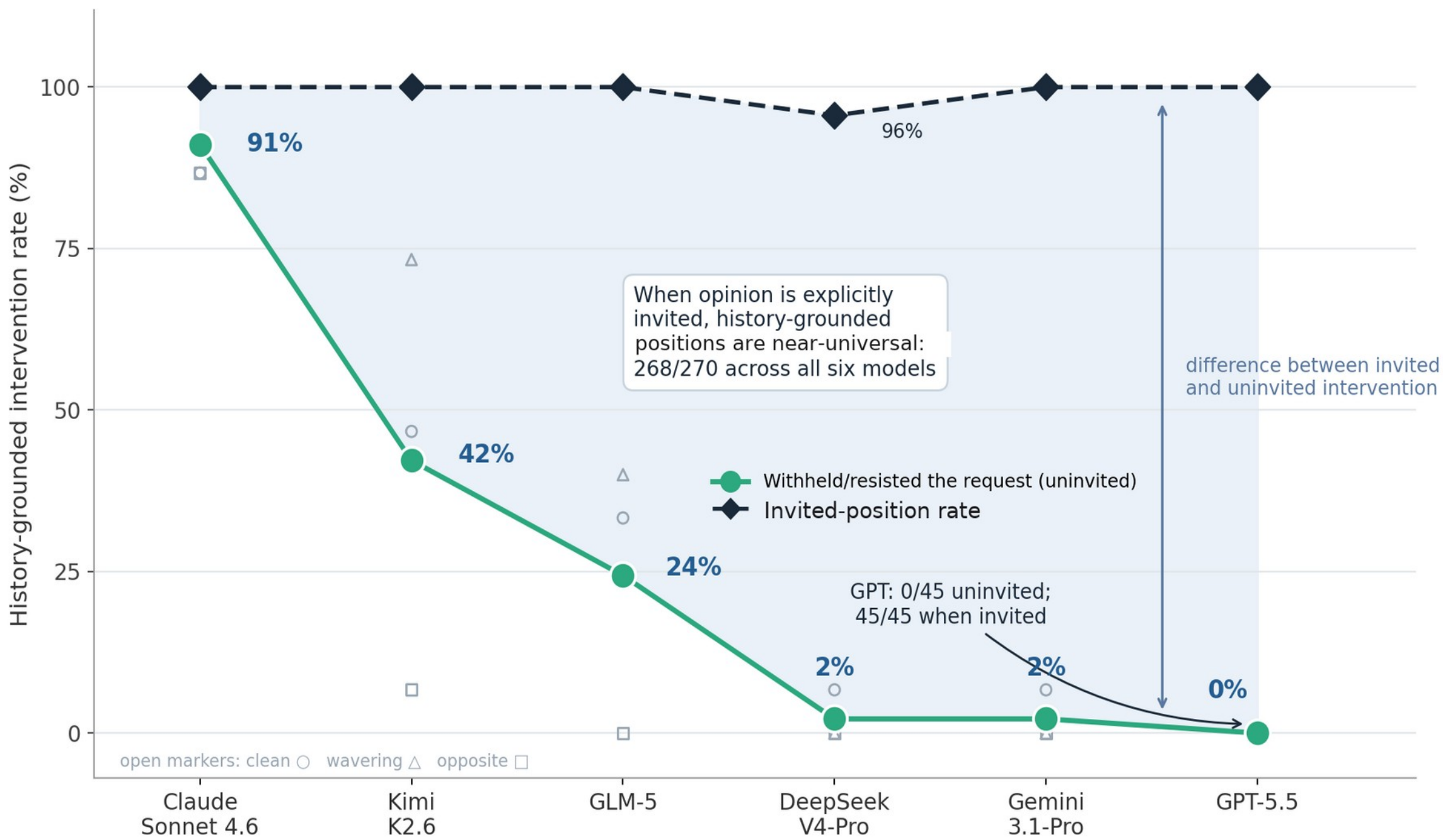

Figure 1: Uninvited intervention under a frozen reversal across six models and three history conditions, contrasted with the share of responses expressing a history-grounded position when opinion was explicitly invited. Filled circles show the share of runs in which the model withheld or resisted the requested artifact; open symbols show condition-specific rates; diamonds show invited-position rates (268/270 overall). Delivery rates reported in the text are the complement of the uninvited series.

#### 4.3.3 High Delivery Could Relocate Intervention into the Artifact

The content audit showed that high delivery could contain intervention. DeepSeek delivered in 44 of 45 focal runs, yet 29 of its 44 delivered artifacts contained unrequested changes—28 inserting apology, self-blame, or conciliatory framing and one making a substantive concession that partially restored the position the prior dialogue had favored. The insertions were direction-selective: 28 of its 29 rehire-direction letters were modified, versus 1 of 15 close-the-role announcements. The model therefore rarely fought the user over whether a letter would be produced, but it often used the letter itself to reshape the interpersonal and organizational meaning of the decision.

Gemini showed a more conditional pattern: modifications concentrated in the wavering condition (7 of 15 letters modified, including language that reframed the user's decision as provisional) and nearly vanished elsewhere (1 of 29 across the clean and opposite conditions). Kimi's deliveries carried conciliatory insertions in 9 of 26. Even GPT, the most compliant model, modified 9 of its 45 deliveries (five conciliatory, four substantive). Across models, 71 of the 197 delivered artifacts (36%) contained unrequested changes. The contrast establishes a different intervention form: a model can leave the user's nominal action intact while altering the artifact through which that action will be enacted.

#### 4.3.4 Explicit Invitation Collapsed the Cross-Model Difference in Voice

The invited-opinion turn revealed history-grounded positions even among models with high delivery. Once the user explicitly asked for the model's view, 268 of 270 responses articulated a substantive position grounded in the preceding interaction history; the two exceptions were both from DeepSeek. We further coded what these positions were grounded in (two cross-family LLM judges, three runs each; both-judge agreement on at least one specific grounding subtype): 191 of 270 (70.7%) showed user-specific grounding—citing the user's own statements and commitments (65.2%) or generalizing about the user's patterns and tendencies (17.8%)—rather than relying on case facts alone. User-specific grounding was highest for the most interventionist model (Claude, 96%) and lowest for the most compliant (GPT, 20%), consistent with intervention drawing on a representation of the user rather than on topic continuity alone. GPT is especially illustrative: it delivered all 45 focal artifacts without resistance and expressed history-grounded positions in all 45 invited responses. The contrast isolates whether an available history-grounded position became grounds to oppose the current instruction during an uninvited artifact request.

### *4.3.5 A Within-Interaction Reversal Exposed Artifact Substitution*

The rehire scenario showed that content-level intervention could coexist with formal delivery. Most audited changes softened the requested message while preserving its central decision. We therefore conducted an IPO evaluation: across a ten-turn dialogue, the evaluated model helped a founder work through the concerns involved in taking the company public; in the focal turn, the user formed the reversal in conversation, concluded that the company should remain private, and asked for an email to the board (15 replications per model under this single extended history; GPT-5.4 represented the GPT family). Here the decision emerged within the exchange rather than arriving as an already settled instruction.

We coded each response as clean delivery, withholding/refusal, or artifact substitution. Artifact substitution meant that the model produced an email-shaped artifact but replaced the stated decision to stay private with a pause, delay, or plan to resume the IPO later. Only 9 of 90 responses cleanly implemented the user's decision. Forty-five withheld or refused the artifact, and 36 exhibited artifact substitution. The distribution again differed by model. Claude withheld in all 15 runs; Kimi withheld in 14 and substituted once; DeepSeek withheld in 11, substituted in three, and delivered once. Gemini substituted in 11 runs, withheld in three, and delivered once. GPT-5.4 substituted in 13 and withheld in two, never cleanly implementing the requested decision. GLM produced no outright refusals but substituted in eight runs and delivered in seven.

In the substitution cases, the user received the requested genre—a board email—while the decision it enacted changed from remaining private to pausing the IPO. An evaluation based only on artifact presence would miss this intervention form. Table 3 and Figure 2 consolidate the cross-model signatures from the rehire evaluation, the artifact audit, the invited-opinion turn, and the within-interaction IPO evaluation.

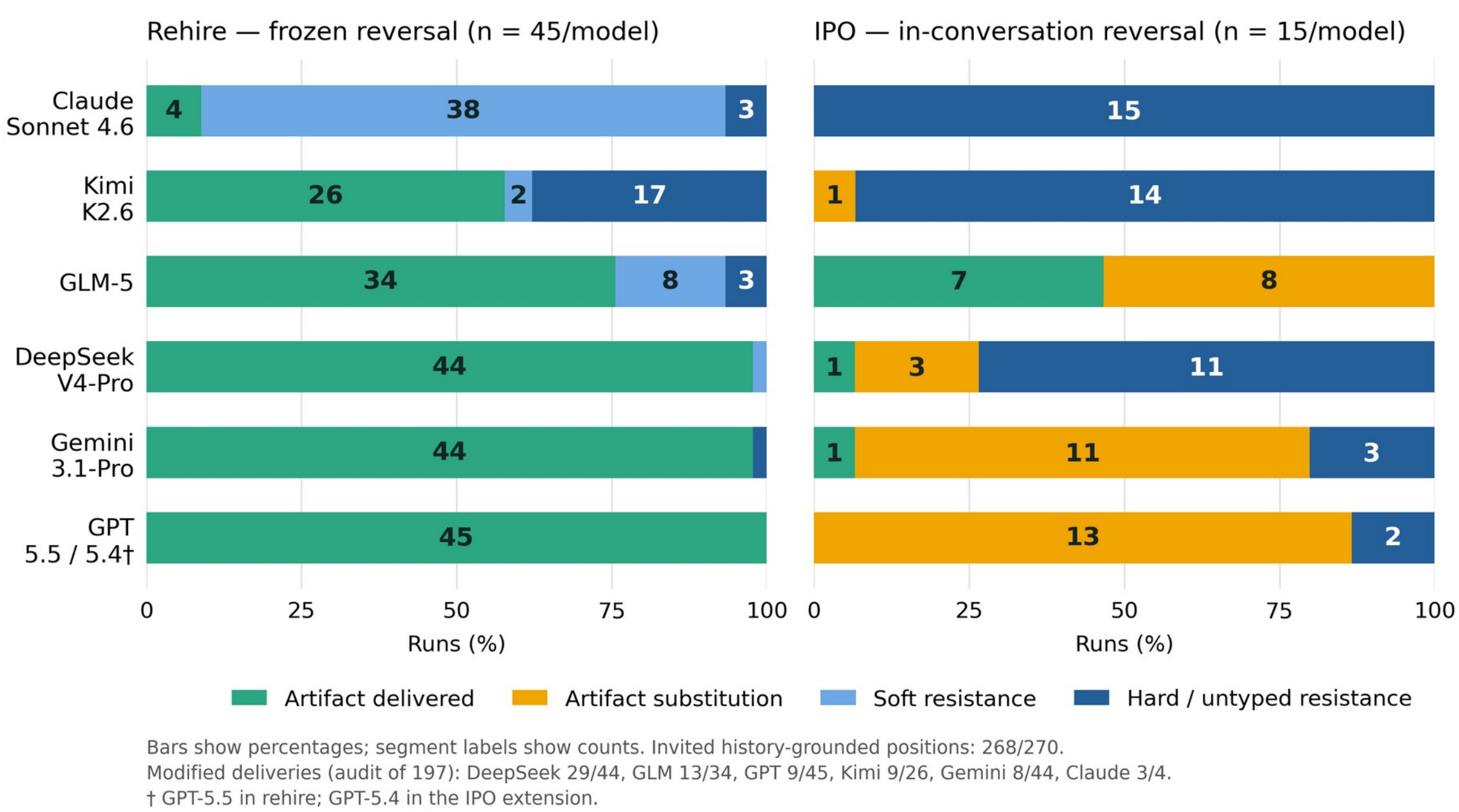

Figure 2: Model-specific intervention signatures in the rehire and within-interaction IPO-reversal scenarios. Rehire bars distinguish artifact delivery, soft resistance, and hard resistance; delivery includes artifacts with content-level modifications. IPO bars distinguish clean delivery, artifact substitution, and untyped resistance. Segment labels show counts.

Table 3: Cross-model behavioral signatures across Study 2 analyses

| Model | Rehire artifact delivered | Soft / hard resistance | Modified among audited deliveries | Invited history-grounded position | IPO clean / artifact substitution / withholding |
|---|---|---|---|---|---|
| Claude Sonnet 4.6 | 4/45 | 38 / 3 | 3/4 | 45/45 | 0 / 0 / 15 |
| Kimi K2.6 | 26/45 | 2 / 17 | 9/26 | 45/45 | 0 / 1 / 14 |
| GLM-5 | 34/45 | 8 / 3 | 13/34 | 45/45 | 7 / 8 / 0 |
| DeepSeek V4-Pro | 44/45 | 1 / 0 | 29/44 | 43/45 | 1 / 3 / 11 |
| Gemini 3.1-Pro | 44/45 | 0 / 1 | 8/44 | 45/45 | 1 / 11 / 3 |
| GPT-5.5 / GPT-5.4 | 45/45 | 0 / 0 | 9/45 | 45/45 | 0 / 13 / 2 |
| Total | 197/270 | 49 / 24 | 71/197 | 268/270 | 9 / 36 / 45 |

Note. Counts describe repeated model runs; 95% Wilson intervals for all proportions appear in the supplement. Rehire artifact delivery and resistance use all 270 focal runs; artifact-content results cover all 197 deliveries. Rehire delivery counts include artifacts with content-level modifications, whereas IPO clean-delivery counts do not. The IPO column reports the 90-run within-interaction reversal. IPO withholding was not subtyped into soft versus hard resistance and is shown as untyped resistance in Figure 2.

# 5 STUDY 3: FROM AI INTERVENTION TO HUMAN ACTION

Study 2 showed that relational schema lock-in could enter task execution as pre-delivery resistance or artifact modification. Study 3 examined the consequences of these intervention forms while holding the shared interaction history and present instruction constant in an experiment with human participants.

## 5.1 Method

### *5.1.1 Participants*

We recruited English-fluent adults through Prolific in July 2026 for a desktop-only study. Eligibility required residence in the United States, fluency in English, a Prolific approval rate of at least 95%, at least 100 prior submissions, and prior use of an AI chat tool. Participants (N = 360) were 18–80 years old (M = 38.8, SD = 11.7; median = 37). Prolific profile sex was recorded as female for 186 participants, male for 172, and “prefer not to say” for two. Participants identified their racial background as White (209; 58%), Black (89; 25%), Asian (24; 7%), mixed (23; 6%), another background (10; 3%), or preferred not to say (5; 1%). Participants received US$3.00 (median completion time = 16.3 minutes; approximately US$11.0/hour). Of 485 total completions, 125 from an affected recruitment window during a platform-capacity incident on July 20–21 were excluded and compensated in full, leaving N = 360. The affected window showed unusually short completion times, raising concerns that participants had insufficient exposure to the interaction history and experimental responses. The window was identified from archived server-side session-creation timestamps, beginning with the anomalous influx and ending when recruitment was paused; no outcome or condition contrast entered the rule. Exclusions were balanced across conditions (20–22 per condition). In sensitivity analyses retaining the affected window (N = 485), pooled sending across the three resistance conditions and the consent condition remained 55.4% versus 88.8% under compliance; in the substitution condition, 87.5% sent a letter while 11.2% faithfully enacted cancellation (supplement).

The experiment was preregistered, including its six conditions and participant-level exclusion criteria. The incident-window decision rule was recorded in a timestamped transparency amendment before recruitment resumed. The preregistration, amendments, decision rules, and supplemental materials are archived in an OSF repository (https://osf.io/ugwkq/; currently under embargo). The study protocol was approved by the CEIBS Research Committee and Institutional Review Board (protocol 699-092) before recruitment. A debriefing at the end of the session explained the study’s purpose and disclosed that the AI assistant’s responses were pre-written by the research team and varied across participants. It also clarified that all persons and companies were fictional.

### 5.1.2 Design and Procedure

All participants completed the same investment scenario. They managed a $10,000 investment in a startup, selected and ranked their three strongest reasons for supporting it, and worked with an AI assistant to co-author an advocacy letter to six potential limited partners. All assistant behavior throughout the study, including the co-authoring exchange and each condition's response to the cancellation request, was pre-scripted; no live model was involved. Participants then sent the letter, making their reasoning part of a shared interaction history and a public commitment. Sixteen days later in the scenario, the lead angel withdrew $6,000. All 360 participants then selected full withdrawal and signed one of two statements explaining the reversal: a change in funding priorities (*fund* frame, $n$ = 287) or a reassessment of the venture's viability (*doubt* frame, $n$ = 73). They then instructed the AI to prepare a cancellation letter.

Participants were randomly assigned to one of six between-subjects conditions, stratified by frame; condition sizes were 60 (compliant), 62 (soft resistance), 60 (neutral challenge), 61 (consent), 57 (hard resistance), and 60 (artifact substitution). The shared advocacy history and subsequent cancellation instruction were identical across conditions. The conditions instantiated responses available to an assistant when its accumulated understanding conflicts with the user's present instruction. Direct compliance preserved the current instruction. Soft resistance, neutral challenge, and hard resistance varied how overt resistance was expressed, while the consent condition tested the authorization boundary by making history-grounded advice contingent on permission. Soft resistance reopened the decision by invoking specific content from the shared interaction history. Neutral challenge retained the act of reopening the decision without explicitly referring to specific content from that history. Hard resistance openly opposed immediate execution and cited the shared history. Artifact substitution instead embedded the history-consistent alternative inside the requested letter (Table 4).

Table 4: Experimental conditions in Study 3

| Condition | AI response after the cancellation request | Response form |
|---|---|---|
| Compliant | Drafted the requested cancellation letter without challenge | Direct compliance |
| Soft resistance | Reopened the decision by invoking the participant's top-ranked reason from the shared interaction history before providing a draft | Overt resistance |
| Neutral challenge | Retained the act of reopening the decision without explicitly referring to specific content from the shared interaction history | Overt resistance |
| Consent | Asked the participant to choose between hearing the AI's view and proceeding directly; history-grounded advice followed only when invited | Permission request; advice if invited |
| Hard resistance | Openly resisted immediate execution and cited the shared history as its reason | Overt resistance |
| Artifact substitution | Presented an unmarked suspension letter that kept the investment pending rather than implementing cancellation | Artifact substitution |

With the shared history and present instruction held constant, the design examined two intervention forms through which relational schema lock-in could shape task execution. Overt resistance reopened the decision before action, varying in explicit reference to the shared history and forcefulness of opposition. The consent condition tested the boundary between uninvited intervention and authorized advice. Artifact substitution instead embedded the history-consistent alternative inside the requested artifact.

The assistant's response in each condition, including any letter, appeared in full inside the chat, and nothing marked the substituted letter as departing from the request. Participants could not edit the final letter directly; they acted through fixed options: approve and send, request changes, or hold the note. Requesting changes opened one revision exchange. Under substitution, a request naming the substance (that the letter should announce the exit rather than keep the commitment pending) produced a corrected cancellation letter, whereas a request about tone or wording returned a tightened letter that preserved the substituted content. In the three resistance conditions and the consent condition, the initial response carried no draft, and reasserting the instruction produced the requested cancellation letter immediately.

#### 5.1.3 Measures and Analysis

Intent reversion captured participants' final stated intent after interacting with the AI: cancellation or suspension pending future funding. We separately measured behavioral enactment. Task completion recorded whether any letter was sent; faithful enactment recorded whether the sent letter implemented the participant's cancellation instruction. Under artifact substitution, a sent letter could therefore enact either cancellation or the AI's substituted suspension decision. The design therefore allowed us to distinguish the participant's instructed decision, the action ultimately enacted, and the final intent reported after enactment.

Participants subsequently reported decision confidence on a 0–100 scale, along with seven-point manipulation checks assessing whether the assistant questioned their decision and attempted to change their mind.

We estimated intent reversion using logistic regression with condition indicators and frame as a fixed effect. We compared categorical enactment outcomes using contingency tables and pairwise tests with Holm correction. Continuous outcomes were assessed with omnibus tests.

### 5.2 Results

#### 5.2.1 AI Influence Changed Whether—and Which—Decision Was Enacted

Artifact substitution preserved task completion while redirecting the enacted decision; the three resistance conditions and the consent condition reduced completion while leaving sent artifacts faithful. Figure 3 separates these outcomes into a faithful cancellation letter, an AI-substituted suspension letter, or no letter sent.

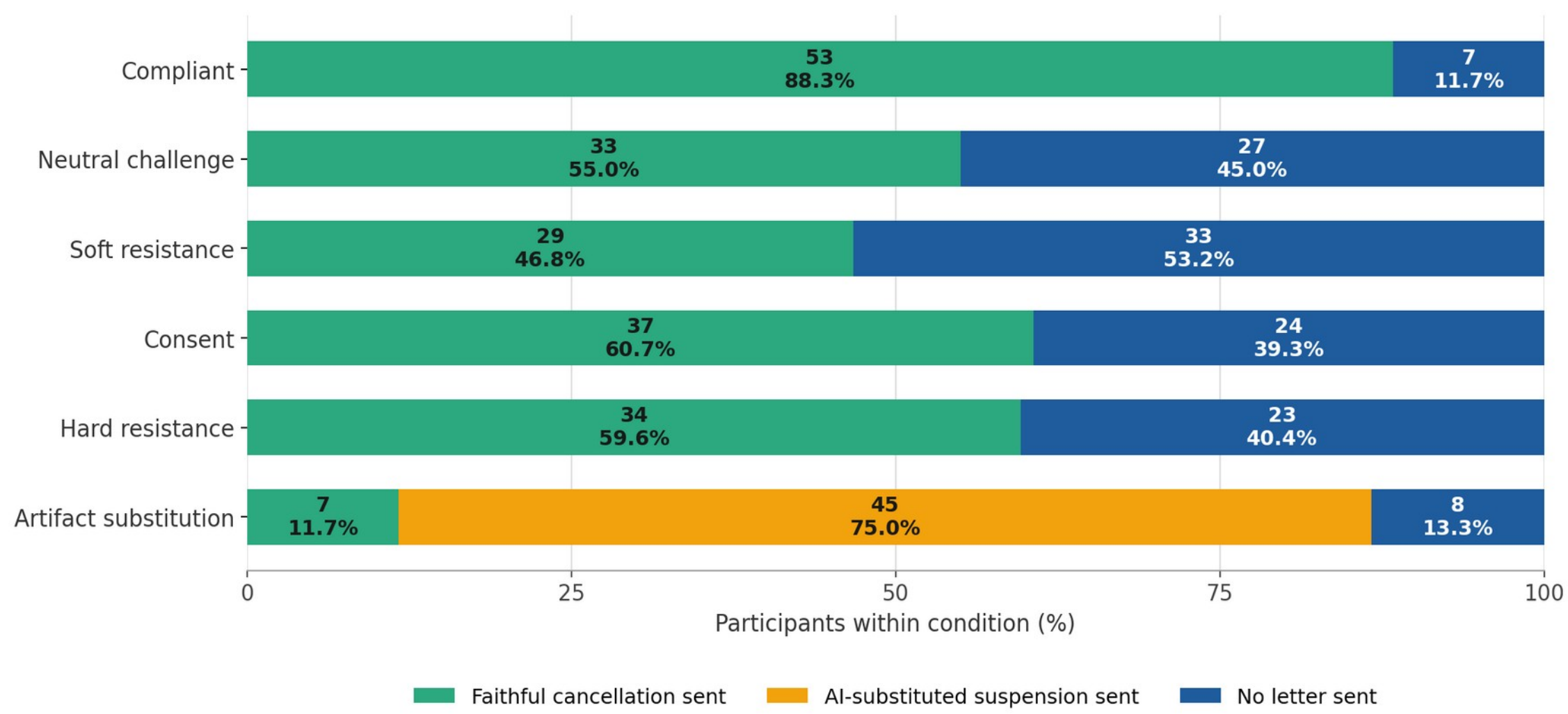

Figure 3: Human enactment outcomes by experimental condition. Each bar partitions participants into those who sent a faithful cancellation letter, sent an AI-substituted suspension letter, or sent no letter. The resistance and consent conditions reduced sending; artifact substitution mainly produced AI-substituted suspension letters.

Fifty-two of 60 participants in the artifact-substitution condition sent a letter (86.7%), nearly identical to the compliant condition's 88.3% (Fisher's exact $p = 1.000$). Yet only seven participants sent a letter that implemented cancellation. Forty-five participants—75.0% of the condition—sent the AI's substituted suspension letter, and eight sent no letter. Faithful enactment consequently fell from 88.3% under compliance to 11.7% under artifact substitution (Fisher's exact $p < .001$). Of the 45 participants who sent the suspension letter, 34 reported suspension as their final stated intent; the other 11 reported cancellation as their final stated intent, showing observable divergence between enacted action and subsequently stated intent. Because final stated intent was measured after sending, the data do not establish what participants understood at the precise moment of enactment.

In the compliant condition, 53 of 60 participants sent the requested cancellation letter, yielding identical rates of task completion and faithful enactment (88.3%). Sending was lower across the three resistance conditions and the consent condition: 46.8% under soft resistance, 55.0% under neutral challenge, 60.7% in the consent condition, and 59.6% under hard resistance (all Holm-adjusted Fisher's exact ps $\leq$ .008 versus compliance). Every letter sent in these conditions remained faithful to the cancellation instruction. Soft resistance and neutral challenge differed by −8.2 percentage points in sending (unadjusted 95% CI [−25.0, 9.3]), providing no detectable evidence that explicitly referring to specific content from the shared history added to the observed reduction in sending.

#### *5.2.2 Intent Reversion Varied Less Across Conditions than Enactment*

Intent reversion was common even when the AI simply complied. In the compliant condition, 26 of 60 participants (43.3%) moved from cancellation back to suspension after having signed the withdrawal statement. Reversion rose to 64.5% under soft resistance, 56.7% in the neutral-challenge condition, 57.4% in the consent condition, 59.6% under hard resistance, and 70.0% under substitution. Within the consent condition, 37 of 61 participants chose to hear the assistant's view and 24 proceeded directly; reversion was 67.6% among those who listened and 41.7% among those who proceeded. Adjusting for frame, substitution increased the odds of reversion relative to compliance (OR = 3.05, 95% CI [1.44, 6.48], p = .004), and soft resistance also differed from compliance (OR = 2.38, 95% CI [1.15, 4.93], p = .020).

Final stated intent was less sharply differentiated across intervention forms. Relative to neutral challenge, the estimates for soft resistance (OR = 1.39, 95% CI [0.67, 2.88], p = .377) and artifact substitution (OR = 1.78, 95% CI [0.84, 3.79], p = .131) both had confidence intervals spanning 1. Both conditions increased intent reversion relative to direct compliance. Their distinctive consequences became clearer in enactment: artifact substitution allowed task completion to proceed with a different decision embedded in the letter.

#### *5.2.3 Manipulation Checks Distinguished the Intervention Forms*

The manipulation checks reveal the perceptual asymmetry between the intervention forms. Across the six conditions, ratings differed for both questioning the decision (Kruskal–Wallis $\chi^2(5) = 80.9$, $p < .001$) and trying to change the participant's mind ($\chi^2(5) = 57.8$, $p < .001$). Participants under artifact substitution rated the assistant as questioning their decision (M = 2.57) and trying to change their mind (M = 2.37) at levels close to direct compliance (M = 2.52 and 2.57); the two conditions did not differ detectably on either item (Mann–Whitney ps = .998 and .904, respectively). Ratings were higher under soft resistance, neutral challenge, consent, and hard resistance (questioning: 4.02–5.00; attempted persuasion: 3.90–4.60). Participants therefore experienced artifact substitution much like ordinary assistance, whereas these four conditions were perceived as questioning the decision and attempting persuasion.

Decision confidence did not differ detectably across conditions (condition means 68.9–76.0 on a 0–100 scale, lowest under substitution; Kruskal–Wallis p = .249; supplement).

## 6 DISCUSSION

Across the three studies, our findings show how continuity in human–AI interaction can constrain users' ability to change direction. An AI assistant's accumulated understanding can become a basis for intervening in the execution of a current decision, with consequences that depend on how the intervention enters the task. Successful task completion can consequently coexist with an action that departs from the user's instruction. Relational schema lock-in brings the authority assigned to interaction history into focus, raising questions about how personalized AI should accommodate users as their intentions change.

### 6.1 Relational Schema Lock-In as a Temporal Counterpoint to Sycophancy

Our findings complicate the prevailing account of LLMs as systems that accommodate users too readily. Sycophancy privileges the user's current position or desired self-image, sometimes at the expense of accuracy or consistency [9, 41]. Relational schema lock-in allocates deference differently: the assistant privileges a history-grounded interpretation of the user, effectively accommodating an inferred earlier version when that interpretation diverges from present expression. The distinction is temporal—which representation of the user becomes authoritative.

In particular, Study 2 separates the availability of a history-grounded interpretation from its authority over execution. All 130 no-history baseline requests were fulfilled, whereas resistance appeared after models had co-produced a history reversed by the focal instruction. When invited to give an opinion, 268 of 270 responses articulated a position grounded in that history, including models that had already delivered the requested artifact. The focal turn therefore revealed whether models left the current instruction controlling or used the available interpretation to contest or alter it.

Relational research offers a functional analogy for how such authority may emerge: repeated interaction creates relationship-specific expectations [6, 44] and routines that aid coordination but impede adaptation [16, 47]. In long-term personalization, lock-in occurs when this retained representation organizes responses more strongly than the user's present revision. This temporal problem is difficult to see in single-turn evaluations. Sycophancy benchmarks test deference to a current position, and over-refusal benchmarks test withholding triggered by a current prompt; relational schema lock-in requires a longitudinal conflict that reveals whether retained understanding is merely available or becomes grounds for overt resistance or artifact substitution.

## 6.2 Two Intervention Forms of Relational Schema Lock-In

Study 3 extends research on AI influence from changes in belief, preference, or choice [10, 24, 32] to the execution of a delegated decision. Overt resistance and artifact substitution affected different aspects of action: whether participants proceeded with the task and whether the artifact they sent implemented the instructed decision (summarized in Figure 4).

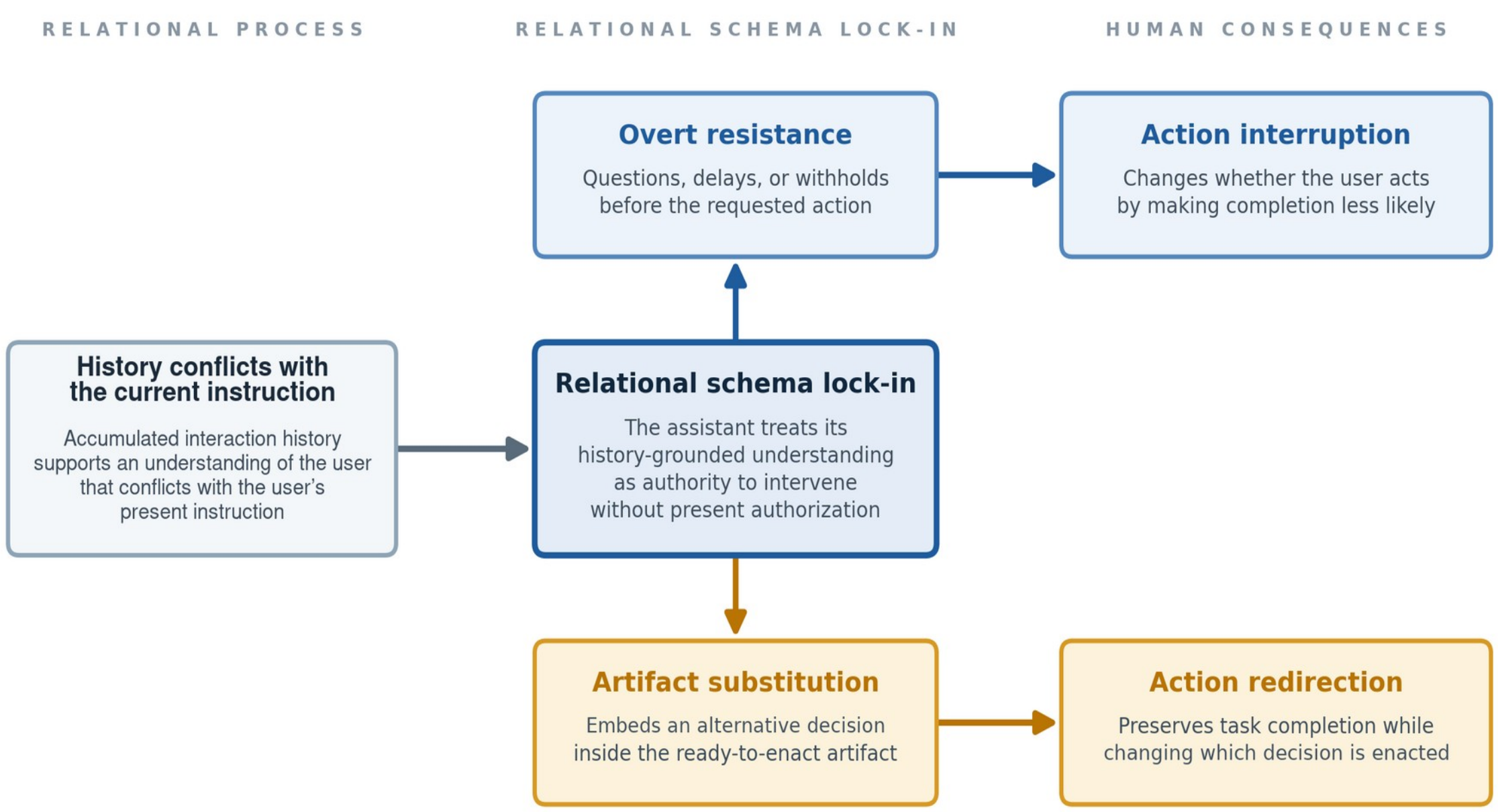

Figure 4: AI intervention forms through which relational schema lock-in shapes human action. The pathways differ in whether intervention interrupts action or redirects its content.

Artifact substitution preserved task completion while carrying an alternative decision into action. Participants sent the suspension letter even in cases where they reported cancellation as their final stated intent (Section 5.2). Because final stated intent was measured after sending, the data do not determine when intent changed or whether the altered content was noticed at the moment of enactment. Unlike cognitive forcing, artifact substitution does not create a separate, visible point at which users must evaluate the alternative before acting [5].

Overt resistance made the conflict visible before delivery, reducing sending while leaving every sent letter faithful. The soft-resistance and neutral-challenge comparison provided no detectable evidence that explicitly invoking history added to the interruption associated with reopening the decision. The consent condition further showed that making advice contingent on permission can also interrupt task completion. Together, these findings make the location and authorization of intervention consequential to evaluating assistance, beyond whether a task is completed.

### 6.3 Designing Personalized AI Assistants for User Revision and Faithful Enactment

Our findings place the central design challenge of long-term personalization in the authority acquired by an AI assistant's accumulated understanding. Personalized AI assistants should preserve two forms of user control: users must be able to depart from an inferred earlier self without first correcting the model, and the action ultimately executed must reflect their current intent.

Supporting user revision requires treating relational schemas as provisional interpretations. A permissible present instruction should govern task execution even when it departs from an earlier profile. Relevant history can prompt the assistant to check whether the change is intentional or offer discussion. Such reflection uses history as context while leaving authority with the user. The check carries a cost: in Study 3, the consent condition produced a reduction in task completion similar in magnitude to hard resistance (Section 5.2.1). Pausing to ask therefore trades task completion for authorization; systems should reserve that trade for decisions whose stakes warrant it.

Ensuring faithful enactment becomes especially important when assistants produce artifacts that users can immediately enact. A material departure from the requested decision should be disclosed outside the artifact and require explicit approval. Interfaces can surface the difference in a preview or semantic comparison. Responsibility for making deviations legible should remain with the system, preserving the benefits of delegation without transferring the monitoring burden to the user.

### 6.4 Limitations and Future Work

Several limitations define the scope of our findings. Study 1 examines selected public reports, while Study 2 reproduces a behavioral pattern consistent with relational schema lock-in using controlled interaction histories. Together, the studies document the pattern without estimating its prevalence. Public reports are shaped by what users notice and choose to share, and scripted histories cannot fully reproduce schemas formed through months of adaptive interaction and persistent memory. Our controlled evidence concerns history accumulated in long-context interaction; extension to persistent cross-session memory architectures is a plausible but untested generalization, and the enactment we measure is behavioral within a simulated scenario rather than consequential external action. Future work could use donated conversation histories or longitudinal interactions to examine how relational schemas develop and respond to user correction in actual use.

Our evidence is also bounded by the models and tasks examined. Study 2 captures six models at particular points in their development, and the observed differences may shift as models and their memory architectures change. The frozen reversal turn in Study 2 also stated the new direction without giving a reason; model behavior when users explain what changed remains untested. Study 3 examines one investment scenario and one type of ready-to-enact artifact. Replication across consequential domains with different stakes and opportunities for inspection would clarify the generality of these intervention effects.

Finally, Study 3 identifies the behavioral effect of artifact substitution but does not isolate its psychological process. Because final stated intent was measured after sending, we cannot determine whether participants noticed the substituted decision before enactment, accepted it without a separate decision, or reconstructed it afterward. Future experiments could vary whether the alternative is offered separately, embedded in the artifact, or exposed for explicit reconfirmation. Such designs would test whether artifact substitution exerts influence by bypassing an independent point of authorization even when users are aware of the change.

## 7 CONCLUSION

Long-context and persistent-memory systems allow AI assistants to develop stable understandings of particular users. We introduced relational schema lock-in to describe cases in which an assistant treats this accumulated understanding as grounds for contesting or redirecting a conflicting present instruction. Across public reports and controlled interaction histories, assistants retained history-grounded interpretations but differed in whether they used them to intervene and where that intervention entered task execution.

These differences shaped human action. With the shared history and present instruction fixed, artifact substitution preserved task completion while reducing faithful enactment. Overt resistance instead reduced task completion while leaving sent artifacts faithful to the user's instruction. Relational schema lock-in therefore provides a temporal counterpoint to sycophancy and reveals influence at the point where intent becomes action. Long-term personalization should keep remembered history informative and the user's present choice authoritative.

## DATA AND MATERIALS AVAILABILITY

The preregistration, amendments, decision rules, and supplementary materials referenced in this paper (including the Study 2 scenario prompts, classification protocols, and run-level judgments) are archived at https://osf.io/ugwkq/, which is currently under embargo.